\documentclass[sigconf]{acmart}

\copyrightyear{2022}
\acmYear{2022}
\setcopyright{acmcopyright}
\acmConference[WWW '22]{Proceedings of the ACM Web Conference 2022}{April 25--29, 2022}{Virtual Event, Lyon, France}
\acmBooktitle{Proceedings of the ACM Web Conference 2022 (WWW '22), April 25--29, 2022, Virtual Event, Lyon, France}
\acmPrice{15.00}
\acmDOI{10.1145/3485447.3512082}
\acmISBN{978-1-4503-9096-5/22/04}

\usepackage{booktabs} % For formal tables
\usepackage{multirow}
\usepackage{subfigure}
\usepackage{verbatim}
\usepackage{CJKutf8}

\usepackage{makecell}
\usepackage[utf8]{inputenc}
\usepackage{flushend}
\begin{document}

\title{FeedRec: News Feed Recommendation with\\ Various User Feedbacks}
\fancyhead{}
%\subtitle{FedCTR}
%\subtitlenote{The full version of the author's guide is available as
%  \texttt{acmart.pdf} document}

\author{Chuhan Wu$^1$, Fangzhao Wu$^{2*}$\authornote{Corresponding Author}, Tao Qi$^1$, Qi Liu$^3$, Xuan Tian$^4$, Jie Li$^5$, Wei He$^5$,\\ Yongfeng Huang$^1$, and Xing Xie$^2$}
\renewcommand{\authors}{Chuhan Wu, Fangzhao Wu, Tao Qi, Qi Liu, Xuan Tian, Jie Li, Wei He, Yongfeng Huang and Xing Xie}

\affiliation{%
  \institution{$^1$Department of Electronic Engineering, Tsinghua University, Beijing 100084, China \\ $^2$Microsoft Research Asia, Beijing 100080, China \quad
  $^3$University of Science and Technology of China, Hefei 230027, China\\
  $^4$Beijing Forestry University, Beijing 100083, China\quad $^5$Microsoft STCA, Beijing 100080, China}
} 
\email{{wuchuhan15, wufangzhao, taoqi.qt}@gmail.com, qiliuql@ustc.edu.cn, tianxuan@bjfu.edu.cn,}
\email{{jieli1,hewe,xingx}@microsoft.com, yfhuang@tsinghua.edu.cn}

% \author{Chuhan Wu}
% \affiliation{%
%     \city{Beijing}
%     \postcode{100084}
%   \state{China}
% }
% \email{wuchuhan15@gmail.com}

% \author{Fangzhao Wu}
% \affiliation{%
%   \institution{Microsoft Research Asia}
%   \city{Beijing}
%   \state{China}
%   \postcode{100080}
% }
% \email{wufangzhao@gmail.com}

% \author{Tao Qi}
% \affiliation{%
%     \city{Beijing}
%     \postcode{100084}
%   \state{China}
% }
% \email{qit16@mails.tsinghua.edu.cn}
% \author{Heyuan Wang}
% \affiliation{%
%   \institution{Microsoft Research Asia}
%   \city{Beijing}
%   \state{China}
%   \postcode{100080}
% }
% \email{heyuanww@163.com}

% \author{Yongfeng Huang}
% \affiliation{%
%     \city{Beijing}
%     \postcode{100084}
%   \state{China}
% }
% \email{yfhuang@tsinghua.edu.cn}

% \author{Xing Xie}
% \affiliation{%
%   \institution{Microsoft Research Asia}
%   \city{Beijing}
%   \state{China}
%   \postcode{100080}
% }
% \email{xing.xie@microsoft.com}

\begin{abstract}

%Personalized news recommendation is critical for news feed platforms to target user interests.
Accurate user interest modeling is important for news recommendation.
Most existing methods for news recommendation rely on implicit feedbacks like click for inferring user interests and model training.
However, click behaviors usually contain heavy noise, and cannot help infer complicated user interest such as dislike.
Besides, the feed recommendation models trained solely on click behaviors cannot optimize other objectives such as user engagement.
In this paper, we present a news feed recommendation method that can exploit various kinds of user feedbacks to enhance both user interest modeling and model training.
We propose a unified user modeling framework to incorporate various explicit and implicit user feedbacks to infer both positive and negative user interests. 
In addition, we propose a strong-to-weak attention network that uses the representations of stronger feedbacks to distill positive and negative user interests from implicit weak feedbacks for accurate user interest modeling.
Besides, we propose a multi-feedback model training framework  to learn an engagement-aware feed recommendation model.
Extensive experiments on a real-world dataset show that our approach can effectively improve the model performance in terms of both news clicks and user engagement.

\end{abstract}




\maketitle

\section{Introduction}

In recent years, online news feed services have gained huge popularity for users to obtain news information from  never-ending feeds on their personal devices~\cite{moniz2018multi}.
However, the huge volume of news articles streaming every day will overwhelm users~\cite{an2019neural}.
Thus, personalized news recommendation is important for news feed services to alleviate information overload and improve the reading experience of users~\cite{li2016context,okura2017embedding,wu2020mind}.

\begin{figure}[!t]
  \centering
    \includegraphics[width=0.48\textwidth]{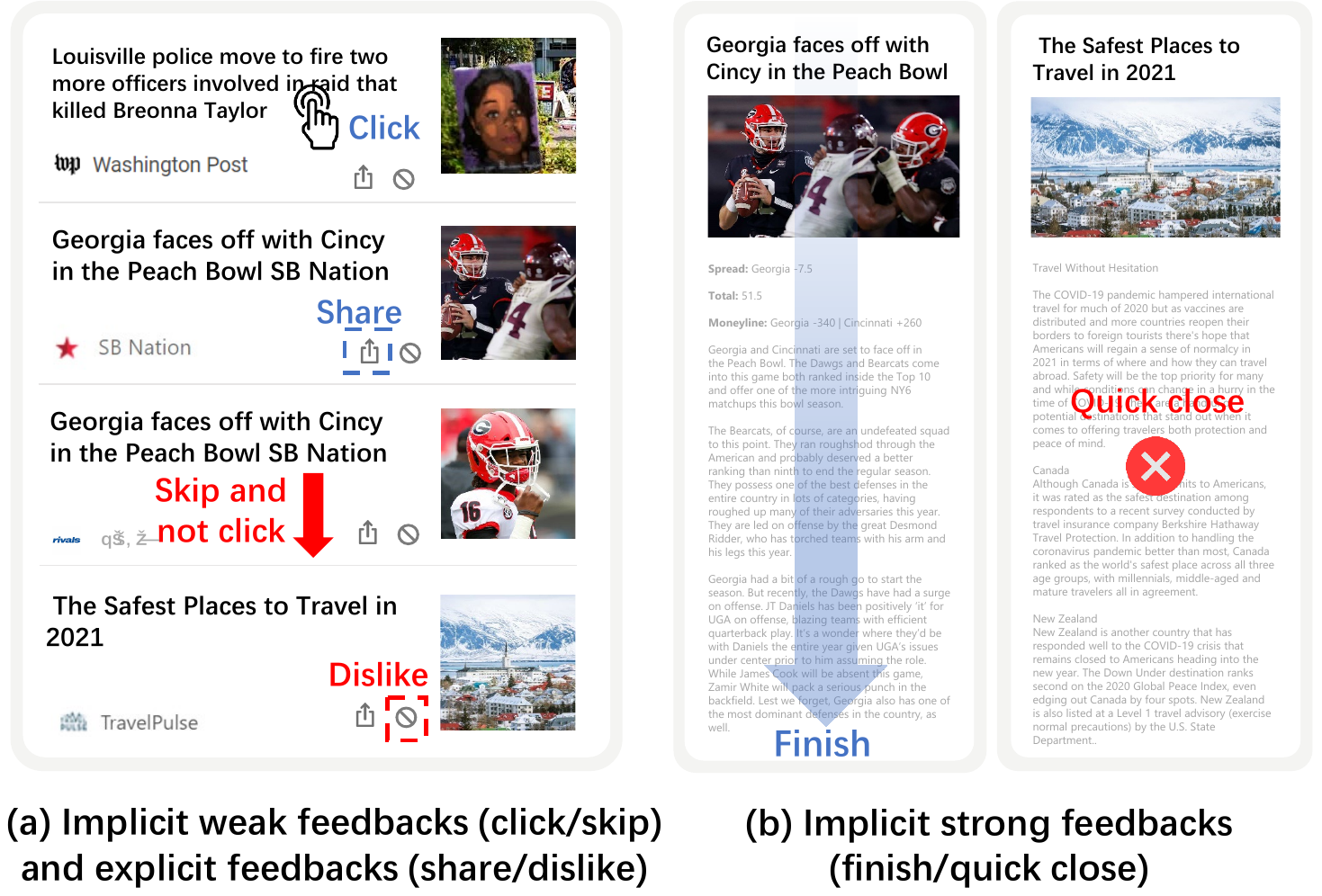}
  \caption{An example of various user feedbacks on a news feed platform.}  \label{exp}
\end{figure}

Most existing news recommendation methods rely on click behaviors of users to infer their interests and train the recommendation model~\cite{okura2017embedding,wang2018dkn,wu2019,wu2019npa,wu2019nrms,wang2020fine}.
For example, Okura et al.~\shortcite{okura2017embedding} proposed to use a GRU network to learn user representations from historical clicked news.
Wang et al.~\shortcite{wang2018dkn} proposed to use a candidate-aware attention network to measure the relevance between clicked news and candidate news when learning user representations.
Wu et al.~\shortcite{wu2019nrms} proposed to use a combination of multi-head self-attention and additive attention networks to learn user representations from clicked news.
All these methods are trained by predicting future news clicks based on the user interests inferred from historical clicked news.
However, click behaviors are implicit feedbacks and usually contain heavy noise~\cite{yi2014beyond,wen2019leveraging}.
For example, users may click a news due to the attraction of a news title but close it quickly if the user is disappointed at the news content~\cite{wu2020cprs}.
In addition, many user interests such as like and dislike cannot be indicated by the implicit click feedbacks, which are actually very important for improving the engagement of users on the news platform.
Thus, it is insufficient to model user interests and train the recommendation model only based on news clicks.

Fortunately, on news feed platforms there are usually various kinds of user feedbacks.
An example is shown in Fig.~\ref{exp}.
Besides the weak implicit feedbacks such as click and skip, there are also explicit feedbacks like share and dislike (Fig.~\ref{exp}(a)) and strong implicit feedbacks like finishing the news article and closing the news webpage quickly after click (Fig.~\ref{exp}(b)).
These feedbacks can provide more comprehensive information for inferring user interests~\cite{tang2016empirical}.
However, it is non-trivial to incorporate the various  feedbacks into news feed recommendation due to several challenges.
First, implicit feedbacks are usually very noisy.
Thus, it is important to distill real positive and negative user interests from the noisy implicit feedbacks.
Second, different feedbacks have very different characteristics, e.g., the intensity of user interests they reflect.
Thus, the model needs to take their differences into consideration.
Third, the feedbacks of a user may have some inherent  relatedness.
For example, a user may quickly close the webpage of a clicked news and then push the dislike button.
Thus, it is important to model the relatedness between feedbacks for better modeling user interests.

In this paper, we present a news feed recommendation approach named \textit{FeedRec}\footnote{Source code is available at https://github.com/wuch15/FeedRec.} that can incorporate various user feedbacks into both user modeling and recommendation model training.
In our method, we propose a unified framework to incorporate various explicit and implicit feedbacks of users, including \textit{click}, \textit{skip}, \textit{share}, \textit{dislike}, \textit{finish}, and \textit{quick close}, to infer both positive and negative interests of users.\footnote{Our approach is a general framework to incorporate various user feedbacks and it is compatible with other types of feedbacks.}
We use a heterogeneous Transformer to capture the relatedness among different kinds of feedbacks, and use several homogeneous Transformers to capture the relations among the same kind of feedbacks.
In addition, we propose a strong-to-weak attention network that uses the representations of stronger feedbacks to distill accurate positive and negative interests from implicit weak feedbacks.
Besides, we propose a multi-feedback model training framework that jointly trains the model using click prediction, finish prediction and dwell time prediction tasks to learn an engagement-aware feed recommendation model.
Extensive experiments on real-world dataset validate that our approach can not only gain more news clicks but also effectively improve user engagement in different aspects.

The contributions of this paper are summarized as follows: 
\begin{itemize}
    \item We propose a unified user modeling framework which can incorporate various explicit and implicit feedbacks to infer both positive and negative user interests.

  \item We propose a strong-to-weak attention network to distill accurate positive and negative  user interests from implicit feedbacks with the guidance of strong feedbacks.

\item We propose a multi-feedback model training framework by jointly training the model in click, finish and dwell time prediction tasks to learn engagement-aware feed recommendation models.

\end{itemize}

\section{Related Work}

User modeling is critical for personalized news recommendation~\cite{li2019survey}. 
Most existing news recommendation approaches model user interests based on historical clicked news~\cite{wu2019,wu2019TANR,zhu2019dan,khattar2018weave,zhang2019dynamic,santosh2020mvl,lee2020news,ge2020graph,hu2020graph,liu2020kred,wu2020sentirec,qi2020privacy,hu2020graphb,qi2021uni,zhang2021amm,tian2021joint,wu2021empowering,wu2021two,wu2021fairness,wu2021uag,yi2021efficient,qi2021hierec,wu2021newsbert,qi2021kim,qi2021pprec}. 
For example, Okura et al.~\shortcite{okura2017embedding} proposed an embedding-based news recommendation method that uses a GRU network to capture user interests from the representations of clicked news.
Wang et al.~\shortcite{wang2018dkn} proposed to use a candidate-aware attention network to learn user representations from clicked news based on their relevance to candidate news.
Wu et al.~\shortcite{wu2019npa} proposed a news recommendation method with personalized attention network that selects informative clicked news for user modeling according to the embeddings of user IDs.
Wu et al.~\shortcite{wu2019nrms} proposed to use multi-head self-attention mechanism to capture the relations between clicked news and use additive attention to select informative news for user modeling.
Wang et al.~\shortcite{wang2020fine} proposed to use  a hierarchical dilated convolution neural network to learn multi-grained features of clicked news for representing users.
These methods only consider the click behaviors of users.
However, click behaviors are usually very noisy for inferring user interests because users may not click news only due to their interests.
In addition, click behaviors cannot reflect many other kinds of user interests such as like or dislike.
Thus, it is insufficient to accurately and comprehensively model user interests with click feedbacks only.

There are only a few news recommendation methods that consider user feedbacks beyond clicks in user modeling~\cite{wu2019neural,wu2020cprs,xie2020deep,ma2021graph,shi2021wg4rec}.
For example,  Gershman et al.~\shortcite{gershman2011news} proposed to represent users by the news they carefully read, rejected, and scrolled.
Yi et al.~\shortcite{yi2014beyond} proposed to use the dwell time of news reading as the weights of clicked news for user modeling.
Wu et al.~\shortcite{wu2020cprs} proposed
a user modeling method based on click preference and reading satisfaction, which uses news clicks and the reading satisfaction derived from dwell time and news content length to model users.
Xie et al.~\shortcite{xie2020deep} proposed to model users' interests by their  click, non-click and dislike feedbacks.
They used click- and dislike-based user representations to  distill positive and negative user interests from non-clicks, respectively.
However, these methods mainly rely on clicked news to model the positive interests of users, which may not be accurate enough due to the heavy noise in click behaviors.
Different from them, our approach can incorporate the various feedbacks of users into user modeling to distill both positive and negative feedbacks, which can capture user interests more comprehensively and accurately.
In addition, our approach jointly trains the model in  various tasks including click prediction, finish prediction and dwell time prediction, which can learn an engagement-aware feed recommendation model.

\section{Methodology}\label{sec:Model}

In this section, we introduce the details of our \textit{FeedRec} approach for news feed recommendation. 
We first introduce its user modeling framework, then  describe the model architecture for news modeling, and finally introduce our multi-feedback model training method.

\subsection{User Modeling}

\begin{figure*}[!t]
  \centering
    \includegraphics[width=0.8\textwidth]{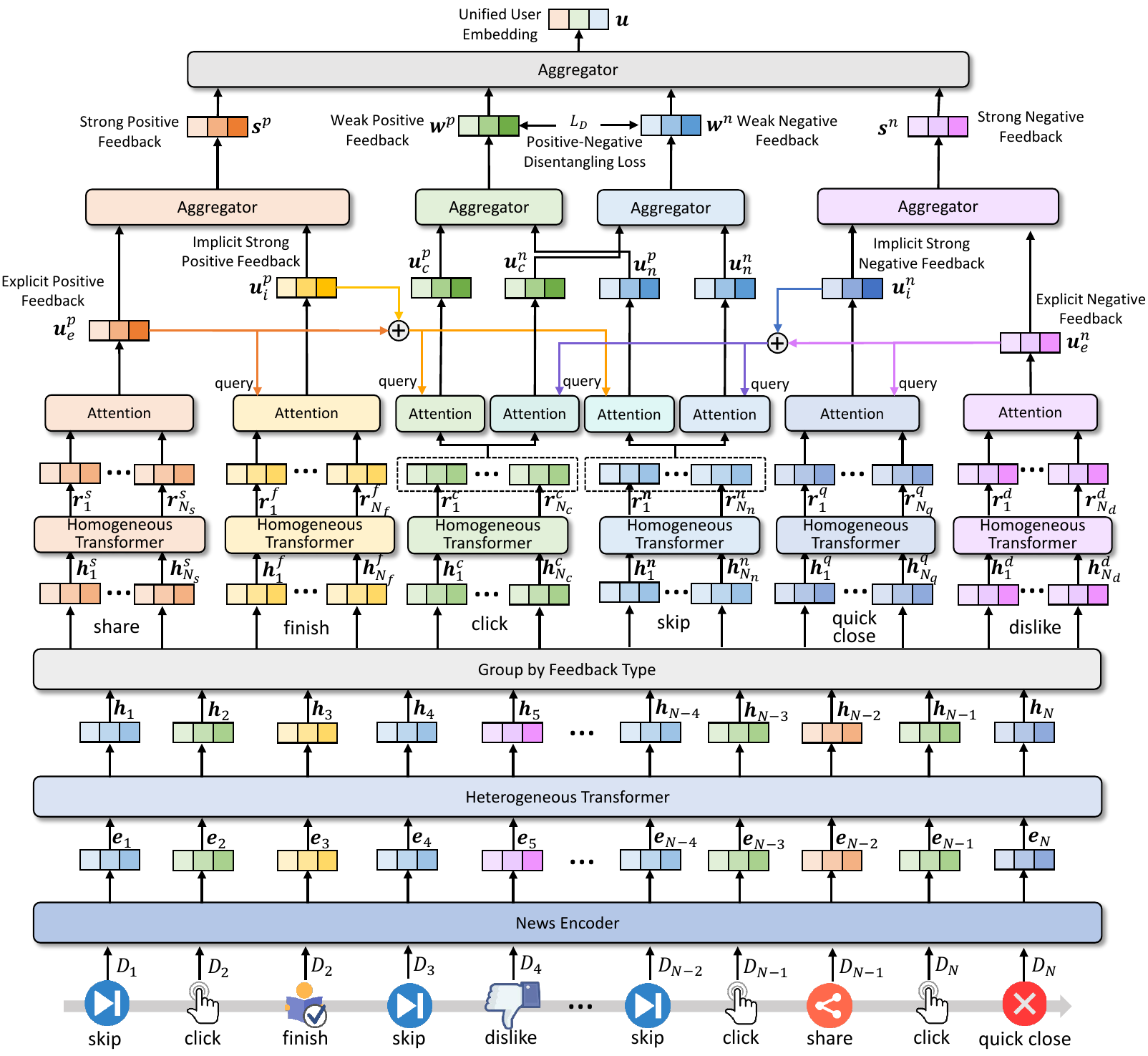}
  \caption{The user modeling framework of our \textit{FeedRec} approach.}  \label{feedmodel}
\end{figure*}

The user modeling framework of our \textit{FeedRec} approach is shown in Fig.~\ref{feedmodel}.
It aims to accurately infer the user preferences for subsequent news feed recommendation by distilling  positive and negative user interests from both explicit and implicit feedbacks it incorporates.
In our approach, we consider six kinds of user feedbacks in total, including \textit{click}, \textit{skip}, \textit{share}, \textit{dislike}, \textit{finish} and \textit{quick close}.
As shown in Fig.~\ref{exp}(a), the \textit{click} feedback is obtained from users' click behaviors on the displayed news articles, which is a commonly used implicit positive feedback for user modeling.
Users can also skip some news without click such as the third news in Fig.~\ref{exp}(a), which is regarded as an implicit negative feedback.
In addition, along with each displayed news, there are  buttons for users to provide explicit feedbacks such as \textit{share} and \textit{dislike}.
For example, the user shares the second news in Fig.~\ref{exp}(a) while reports a dislike of the fourth news.
Besides, there are also implicit feedback stronger than \textit{click} and \textit{skip}.
For example, as shown in Fig.~\ref{exp}(b), after a user clicking a news, this user may finish reading this news  (including watching the embedded video), which usually indicates a positive interest.
However, the user may also take a quick read  after click  for only a few seconds and then close the news webpage, which is an indication of dissatisfaction.
We use the news reading behavior with dwell time shorter than $T$ seconds to construct this kind of feedback.

Next, we introduce the architecture of our user modeling framework.
We first use a shared news encoder to obtain the embedding of each feedback and its associated news article.
We denote the feedback sequence as $[D_1, D_2, ..., D_N]$, where $N$ is the sequence length.\footnote{Some feedbacks may occur on the same news, e.g., finishing after clicking.}
It is converted into a feedback embedding sequence, which is denoted as $\mathbf{E}=[\mathbf{e}_1, \mathbf{e}_2, ..., \mathbf{e}_N]$.

Next, we apply a heterogeneous feedback Transformer~\cite{vaswani2017attention} to the feedback embedding sequence to capture the relations between different feedbacks.
The feedbacks from the same user may have some inherent relatedness~\cite{xie2020deep}.
For example, the \textit{finish} and \textit{quick close} feedbacks usually appear after clicks.
In addition, some skips may also have correlations to the previous clicks because a user may only choose to read a few news on similar topics~\cite{li2011scene}.
For example, in Fig.~\ref{exp}(a) the user clicks and shares the second news while skips the third news, which may be because both of them are about the same football team. 
Thus, we use a heterogeneous feedback Transformer to capture the relations among various kinds of feedbacks in a feedback sequence.
It receives the feedback embedding sequence $\mathbf{E}$ as the input, and outputs a hidden feedback representation sequence $\mathbf{H}=[\mathbf{h}_1, \mathbf{h}_2, ..., \mathbf{h}_N]$.
To help the subsequent user modeling process that  separately models different kinds of feedbacks, we group the hidden feedback representations by their types.
We denote the embedding sequences of \textit{share}, \textit{finish}, \textit{click}, \textit{skip}, \textit{quick close} and \textit{dislike} feedbacks respectively as $\mathbf{H}^s=[\mathbf{h}^s_1, \mathbf{h}^s_2, ... \mathbf{h}^s_{N_s}]$, 
$\mathbf{H}^f=[\mathbf{h}^f_1, \mathbf{h}^f_2, ... \mathbf{h}^f_{N_f}]$,
$\mathbf{H}^c=[\mathbf{h}^c_1, \mathbf{h}^c_2, ... \mathbf{h}^c_{N_c}]$,
$\mathbf{H}^n=[\mathbf{h}^n_1, \mathbf{h}^n_2, ... \mathbf{h}^n_{N_n}]$,
$\mathbf{H}^q=[\mathbf{h}^q_1, \mathbf{h}^q_2, ... \mathbf{h}^q_{N_q}]$ and $\mathbf{H}^d=[\mathbf{h}^d_1, \mathbf{h}^d_2, ... \mathbf{h}^d_{N_d}]$, where 
$N_s$, $N_f$, $N_c$, $N_n$, $N_q$ and $N_d$ are the numbers of the corresponding feedbacks.

Following is a homogeneous feedback Transformer, which is applied to each kind of feedbacks to learn feedback-specific representations.
Different kinds of feedbacks usually have very different characteristics.
For example, \textit{click} and \textit{skip} feedbacks are usually abundant but noisy, while \textit{share} and \textit{dislike} feedbacks are strong but sparse.
Thus, they may need to be handled differently.
In addition, the relations between the same kind of feedbacks are also important for user interest modeling~\cite{xie2020deep}.
For example, researchers have found that modeling the interactions between clicked news can help better infer user interests~\cite{wu2019nrms}. 
Since the heterogeneous Transformer may not focus on capturing the relatedness between homogeneous feedback, we apply independent Transformers to each kind of feedbacks to learn feedback-specific representations for them and meanwhile capture the relations among homogeneous feedbacks.
We denote the feedback-specific representation sequences of \textit{share}, \textit{finish}, \textit{click}, \textit{skip}, \textit{quick close} and \textit{dislike} as  $\mathbf{R}^s=[\mathbf{r}^s_1, \mathbf{r}^s_2, ... \mathbf{r}^s_{N_s}]$,
$\mathbf{R}^f=[\mathbf{r}^f_1, \mathbf{r}^f_2, ... \mathbf{r}^f_{N_f}]$,
$\mathbf{R}^c=[\mathbf{r}^c_1, \mathbf{r}^c_2, ... \mathbf{r}^c_{N_c}]$,
$\mathbf{R}^n=[\mathbf{r}^n_1, \mathbf{r}^n_2, ... \mathbf{r}^n_{N_n}]$,
$\mathbf{R}^q=[\mathbf{r}^q_1, \mathbf{r}^q_2, ... \mathbf{r}^q_{N_q}]$ and $\mathbf{R}^d=[\mathbf{r}^d_1, \mathbf{r}^d_2, ... \mathbf{r}^d_{N_d}]$, respectively.

Based on the representation sequences of each kind of feedbacks, we then propose a strong-to-weak attention network to distill accurate positive and negative interests from implicit weak feedbacks (e.g., \textit{clicks}) based on their relevance to stronger feedbacks (e.g., \textit{share} and \textit{finish}).
Since explicit feedbacks like \textit{share} and \textit{dislike} are usually reliable, we can directly regard them as pure positive and negative feedbacks, respectively.
We apply two separate attention networks~\cite{yang2016hierarchical} to them to learn an explicit positive feedback representation $\mathbf{u}^p_e$ and an explicit negative feedback representation $\mathbf{u}^n_e$, which are formulated as follows:
\begin{equation}
\begin{aligned}
     \alpha^p_k &=\frac{\exp(\mathbf{q}^s\cdot  \mathbf{r}^s_k)}{\sum_{j=1}^{N_s}\exp(\mathbf{q}^s\cdot  \mathbf{r}^s_j)}, ~~~~
     \mathbf{u}^p_e =\sum_{k=1}^{N_s} \alpha^p_k \mathbf{r}^s_k, \\
\end{aligned}
\end{equation}
\begin{equation}
\begin{aligned}
     \alpha^n_k &=\frac{\exp(\mathbf{q}^d\cdot  \mathbf{r}^d_k)}{\sum_{j=1}^{N_d}\exp(\mathbf{q}^d\cdot  \mathbf{r}^d_j)}, ~~~~
     \mathbf{u}^n_e =\sum_{k=1}^{N_d} \alpha^n_k \mathbf{r}^d_k.
\end{aligned}
\end{equation}
Next, we use the explicit positive feedback representation $\mathbf{u}^p_e$ to select informative \textit{finish} feedbacks and build a representation  $\mathbf{u}^p_i$ of implicit strong positive feedback, which is formulated as follows:
\begin{equation}
\begin{aligned}
     \beta^p_k &=\frac{\exp( \mathbf{u}^p_e\cdot  \mathbf{r}^f_k)}{\sum_{j=1}^{N_f}\exp(\mathbf{u}^p_e\cdot  \mathbf{r}^f_j)},~~~~
     \mathbf{u}^p_i =\sum_{k=1}^{N_f} \beta^p_k \mathbf{r}^f_k.
\end{aligned}
\end{equation}
The implicit strong negative feedback $\mathbf{u}^n_i$ is computed in a similar way from the representations of \textit{quick close} feedbacks as follows:
\begin{equation}
\begin{aligned}
     \beta^n_k &=\frac{\exp( \mathbf{u}^n_e\cdot  \mathbf{r}^q_k)}{\sum_{j=1}^{N_q}\exp(\mathbf{u}^q_e\cdot  \mathbf{r}^q_j)},~~~~
     \mathbf{u}^n_i =\sum_{k=1}^{N_q} \beta^n_k \mathbf{r}^q_k.
\end{aligned}
\end{equation}
\textit{Click} and \textit{skip} feedbacks are usually noisy for inferring positive and negative interests~\cite{xie2020deep,wu2020cprs}.
This is because clicks do not necessarily mean like or satisfaction, and those seen but skipped news may also be relevant to user interests.
Thus, we need to distill the real positive and negative user interests from them.
To address this problem, we select \textit{click} and \textit{skip} feedbacks based on their relevance to strong feedbacks for learning positive and negative user interest representations.
We use the summation of $\mathbf{u}^p_e$ and $\mathbf{u}^p_i$ as the attention query for distilling the click-based and skip-based weak positive interests (denoted as $\mathbf{u}^p_c$ and $\mathbf{u}^p_n$), which are computed as follows:
\begin{equation}
\begin{aligned}
     \gamma^p_k &=\frac{\exp[ (\mathbf{u}^p_e+\mathbf{u}^p_i)\cdot  \mathbf{r}^c_k]}{\sum_{j=1}^{N_c}\exp[\mathbf{u}^p_e+\mathbf{u}^p_i)\cdot  \mathbf{r}^c_j]},~~~~
     \mathbf{u}^p_c  =\sum_{k=1}^{N_c} \gamma^p_k \mathbf{r}^c_k,
\end{aligned}
\end{equation}
\begin{equation}
\begin{aligned}
     \gamma^n_k &=\frac{\exp[ (\mathbf{u}^p_e+\mathbf{u}^p_i)\cdot  \mathbf{r}^n_k]}{\sum_{j=1}^{N_n}\exp[\mathbf{u}^p_e+\mathbf{u}^p_i)\cdot  \mathbf{r}^n_j]},~~~~
     \mathbf{u}^p_n  =\sum_{k=1}^{N_n} \gamma^n_k \mathbf{r}^n_k.
\end{aligned}
\end{equation}
The click- and skip-based weak negative feedbacks (denoted as $\mathbf{u}^n_c$ and $\mathbf{u}^n_n$) are computed similarly by using  $\mathbf{u}^n_e+\mathbf{u}^n_i$ as the attention query.
In this way, we can distill accurate positive and negative user interest information from the noisy feedbacks.

The last one is feedback aggregation.
It aims to aggregate different kinds of feedbacks into summarized representations by considering their different importance and functions.
We first aggregate the explicit positive feedback $\mathbf{u}^p_e$ and implicit strong positive feedback  $\mathbf{u}^p_i$ into a unified strong positive feedback representation $\mathbf{s}^p$, which is formulated as follows:
\begin{equation}
    \delta^p=\sigma(\mathbf{v}^{p}\cdot [\mathbf{u}^p_e;\mathbf{u}^p_i]), 
    \mathbf{s}^p=\delta^p\mathbf{u}^p_e+(1-\delta^p)\mathbf{u}^p_i,
\end{equation}
where $\sigma$ is the sigmoid function, $\mathbf{v}^{p}$ is a learnable vector.
In a similar way, we aggregate the explicit negative feedback $\mathbf{u}^n_e$ and implicit strong negative feedback  $\mathbf{u}^n_i$ into a unified strong negative feedback representation $\mathbf{u}^n$ as follows:
\begin{equation}
    \delta^n=\sigma(\mathbf{v}^{n}\cdot [\mathbf{u}^n_e;\mathbf{u}^n_i]), 
    \mathbf{s}^n=\delta^n\mathbf{u}^n_e+(1-\delta^n)\mathbf{u}^n_i,
\end{equation}
where $\mathbf{v}^{n}$ are parameters.
Similarly, we aggregate the click-based and skip-based positive feedbacks ($\mathbf{u}^p_c$ and $\mathbf{u}^p_n$) into a weak positive feedback representation $\mathbf{w}^p$, and aggregate  $\mathbf{u}^n_c$ and $\mathbf{u}^n_n$ into a weak negative feedback representation $\mathbf{w}^n$.
We finally aggregate the four kinds of feedbacks, i.e., $\mathbf{s}^p$, $\mathbf{w}^p$, $\mathbf{w}^n$ and $\mathbf{s}^n$ into a unified user embedding $\mathbf{u}$, which is formulated as follows:
\begin{equation}
\mathbf{u}=s^p\mathbf{s}^p+w^p\mathbf{w}^p+s^n\mathbf{s}^n+w^n\mathbf{w}^n,
\end{equation}
where $s^p$, $w^p$, $s^n$, $w^n$ are learnable parameters.

\subsection{News Modeling}

\begin{figure}[!t]
  \centering
    \includegraphics[width=0.3\textwidth]{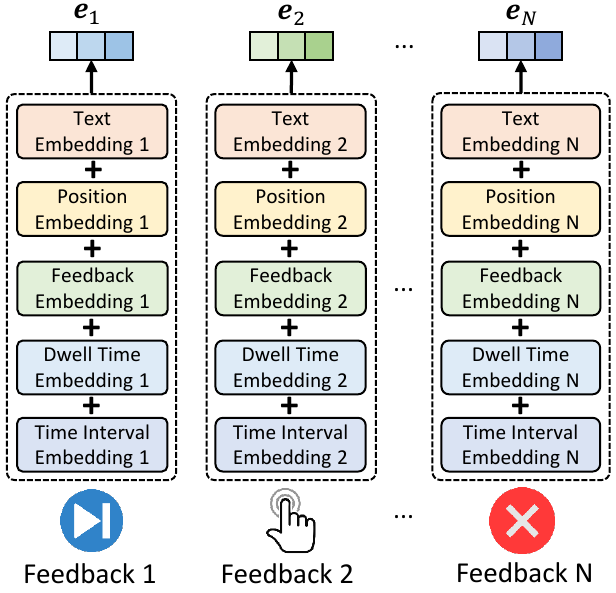}
  \caption{The architecture of the news encoder.}  \label{newsencoder}
\end{figure}

In this section, we briefly introduce the details of news encoder in our approach.
The architecture of the news encoder is shown in Fig.~\ref{newsencoder}.
For each feedback on news, we compute five kinds of embeddings for it.
The first one is text embedding, which is computed from news title through a Transformer~\cite{vaswani2017attention} network to capture the semantic information of news.
The second one is position embedding, which aims to encode the positional information of feedback.
The third one is feedback embedding, which encodes the type of feedback to help better distinguish different kinds of feedbacks.\footnote{This embedding is deactivated when encoding candidate news.}
The fourth one is dwell time embedding, which aims to encode use engagement information (we use the user-specific dwell time).
We use a quantization function $\tilde{t}=\lfloor \log_2(t+1)\rfloor$ to convert the real-valued dwell time $t$ into a discrete value $\tilde{t}$ for building the embedding table. 
The last one is time interval embedding, which aims to better capture the relatedness between adjacent feedbacks.
We use the same quantization function to convert the
time interval between the current and previous feedbacks into a discrete variable for embedding.
These embeddings are added together into a unified news embedding for subsequent user modeling and model training.

\subsection{Multi-feedback Model Training}

\begin{figure}[!t]
  \centering
    \includegraphics[width=0.36\textwidth]{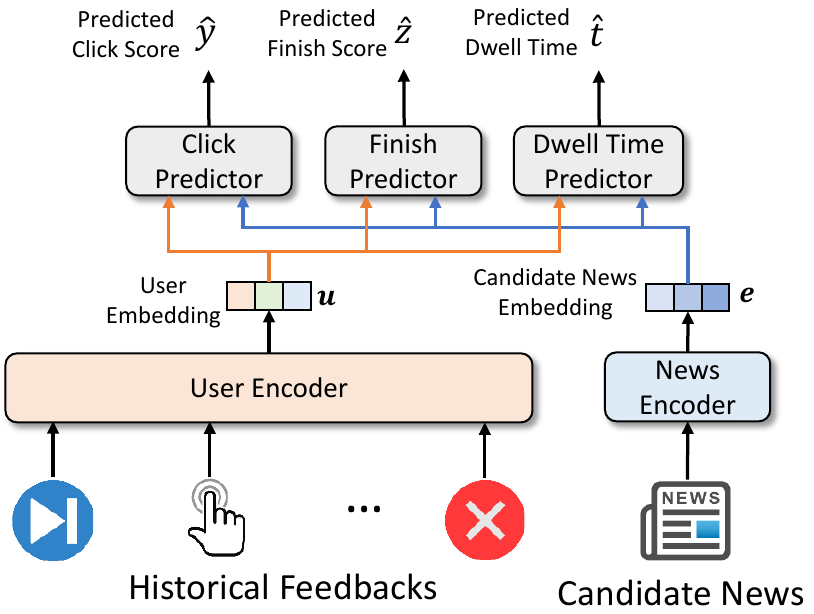}
  \caption{The multi-feedback model training framework.}  \label{feedtrain}
\end{figure}

In this section, we introduce the multi-feedback framework in our approach.
Existing news recommendation methods mainly rely on the click signals to train the recommendation model.
However, there are usually some gaps between news clicks and user engagement or satisfaction, because users may leave the news page quickly if they are not satisfied with the quality of news content.
Thus, we propose to jointly train the model in three tasks, including click prediction, finish prediction and dwell time prediction, to encode both click and user engagement information.
The model training framework is shown in Fig.~\ref{feedtrain}.
We use the user encoder to learn a user embedding $\mathbf{u}$ from the feedback sequence and use the news encoder to encode the candidate news into its embedding $\mathbf{e}$.
We denote the predicted click, finish and dwell time scores of this pair of user and candidate news as $\hat{y}$, $\hat{z}$ and $\hat{t}$ respectively, which are computed as follows:
\begin{equation}
\begin{aligned}
    \hat{y}&=\mathbf{u}\cdot \mathbf{e},\\
\hat{z}&=\mathbf{u} \cdot (\mathbf{W}_z \mathbf{e}),\\
\hat{t} &=\max[0,\mathbf{u}\cdot (\mathbf{W}_t \mathbf{e})],
\end{aligned}
\end{equation}
 where $\mathbf{W}_z$ and $\mathbf{W}_t$ are learnable parameters.

Following~\cite{wu2019nrms}, we use negative sampling techniques to construct training samples.
For each clicked news, we sample $K$ skipped news displayed on the same page, and jointly predict the three kinds of scores for these $K+1$ news.
The  click, finish and dwell time prediction losses on a sample are formulated as follows:
\begin{equation} 
\begin{aligned}
   \mathcal{L}_R&=-\log[\frac{\exp(\hat{y}^+)}{\exp(\hat{y}^+)+\sum_{i=1}^K\exp(\hat{y}^-_i)}],\\
   \mathcal{L}_F&=-z^+\log[\sigma(\hat{z}^+)]-(1-z^+)\log[1-\sigma(\hat{z}^+)],\\
   \mathcal{L}_T&=|t^+-\hat{t}^+|,
\end{aligned}
\end{equation}
where $\hat{y}^+$ and  $\hat{y}^-_i$ are the predicted click scores for a clicked news and its associated $i$-th skipped news.
$\hat{z}^+$, $z^+$, $\hat{t}^+_i$ and $t^+_i$  stand for the predicted finish label, real finish label,  predicted dwell time, and real dwell time of a clicked news, respectively.\footnote{We use the log function to transform the raw dwell time and then normalize it.}

Besides, since we expect the weak positive feedback to be different from the weak negative feedback, we propose a positive-negative disentangling loss $\mathcal{L}_d$ to help distill more accurate positive and negative user interests by regularizing $\mathbf{w}^p$ and $\mathbf{w}^n$ as follows:
\begin{equation}
\mathcal{L}_D=\frac{\mathbf{w}^p\cdot \mathbf{w}^n}{||\mathbf{w}^p||\times||\mathbf{w}^n||},
\end{equation}
where $||\cdot||$ means the L2-norm.
The final unified loss $\mathcal{L}$  is a weighted summation of  four loss functions, which is formulated as follows:
\begin{equation} 
   \mathcal{L}=\mathcal{L}_R+\alpha \mathcal{L}_F+\beta \mathcal{L}_T+\gamma \mathcal{L}_D,
\end{equation}
where $\alpha$, $\beta$ and $\gamma$ are loss coefficients that control the relative importance of the corresponding loss functions.

\section{Experiments}\label{sec:Experiments}

\subsection{Dataset and Experimental Settings}

\begin{figure}[!t]
	\centering
	\includegraphics[width=0.4\textwidth]{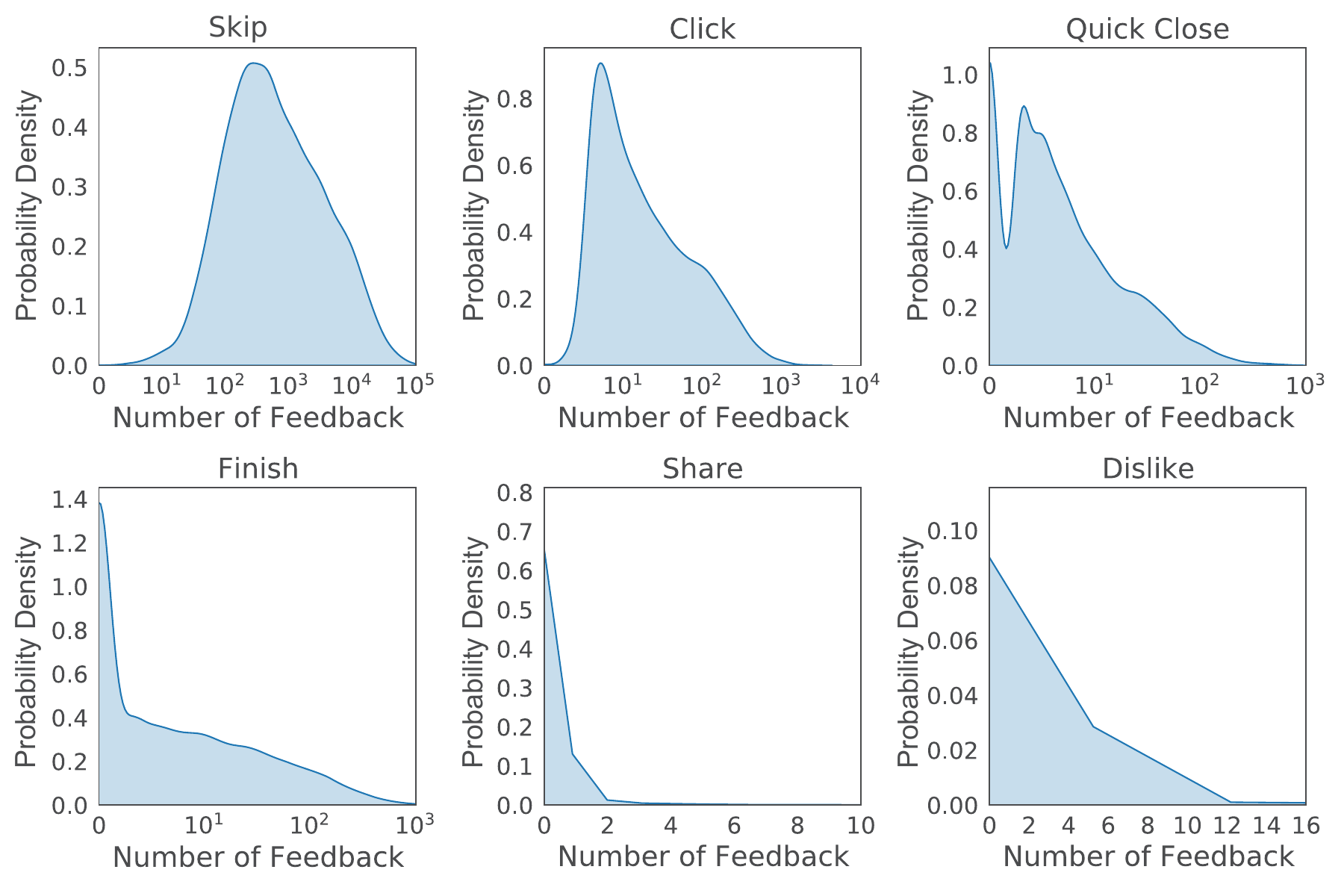}
\caption{Distribution of different kinds of feedbacks.}\label{fig.feeddis}  
\end{figure}

\begin{figure}[!t]
	\centering
	\includegraphics[width=0.24\textwidth]{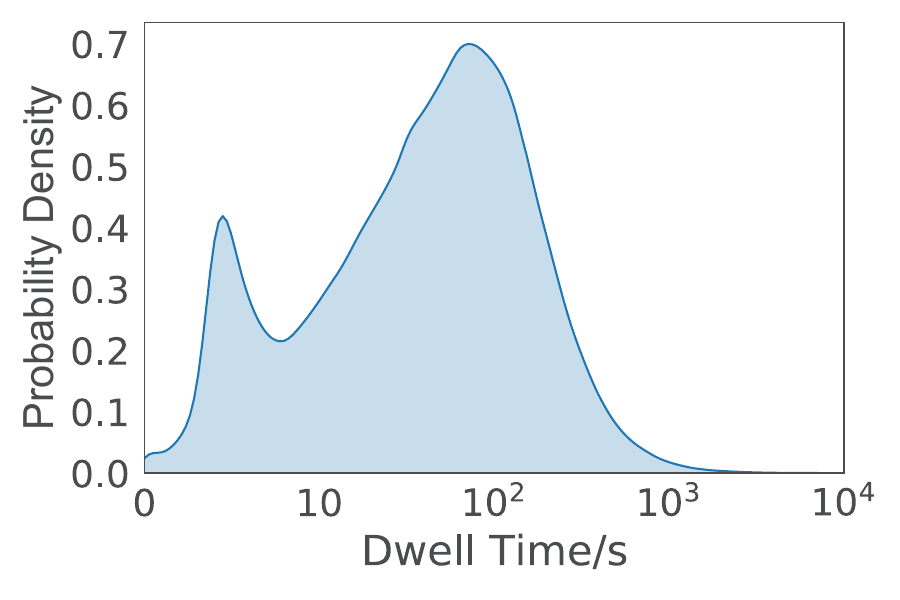}
\caption{Dwell time distribution of the dataset.}\label{fig.dtdis}  
\end{figure}

In our experiments, since there is no off-the-shelf dataset for news recommendation that contains multiple kinds of user feedbacks, we constructed one by ourselves from a commercial news feed App.
The dataset contains the behavior logs of 10,000 users in about one month, i.e., from Sep. 1st, 2020 to Oct. 2nd, 2020.
The logs in the last week were used for test, and the rest ones were used for training and validation (rest logs on the last day).
The statistics of this dataset is shown in Table~\ref{dataset}.
We can see that explicit feedbacks like \textit{share} and \textit{dislike} are relatively sparse, while implicit feedbacks are much richer.
The distributions of the number of each kind of feedback provided by a user are shown in Fig.~\ref{fig.feeddis}.
We can find that the number of \textit{skip} feedbacks is  approximately log-normal, while the numbers of other kinds of feedbacks obey long-tail distributions.
Since \textit{skip} feedbacks are dominant in our dataset, we only randomly sample 10\% of skips to reduce the length of input sequence.
We also show the distribution of dwell time in our dataset in Fig.~\ref{fig.dtdis}.
We find an interesting phenomenon is that the distribution has two peaks, one of which approximately appears between 0 and 10 seconds.
This may be because users are sometimes disappointed at the news content and quickly close the webpage.
Thus, we accordingly set the  dwell time threshold $T$ to 10 seconds to construct the \textit{quick close} feedbacks, and we will discuss the influence of $T$ in the hyperparameter analysis section.

\begin{table}[t]
\centering
\caption{Detailed statistics of the datasets.}\label{dataset} 
%\resizebox{0.48\textwidth}{!}{ 
\begin{tabular}{lrlr}
\Xhline{1.5pt}
\# user        & 10,000     & \# news         & 590,485 \\
\# impression  & 351,581    & \# click        & 493,266 \\
\# skip   & 25,986,877 & \# share        & 2,764   \\
\# dislike     & 17,073     & \# finish       & 234,759 \\
\# quick close & 108,396    & avg. dwell time & 83.90s  \\\Xhline{1.5pt}
\end{tabular} 
%}   

\end{table}

In our experiments, we followed the same settings in~\cite{wu2019nrms} to generate the 256-dim text embeddings, and the dimensions of other embeddings in the news encoder were also 256.
The Transformers in the user modeling part had 16 heads, and the output dimension of each head was 16.
The feedback type, position, dwell time, and time interval  embeddings are randomly initialized.
The optimizer for model training was Adam~\cite{kingma2014adam}, and the learning rate was 1e-4.
The negative sampling ratio was 4.
The batch size was 32.
The dropout~\cite{srivastava2014dropout} ratio was set to 0.2.
These hyperparameters were tuned on the validation sets.
We used AUC, MRR, nDCG@5 and HR@5 to measure the click-based model performance.
In addition, we used several metrics to measure the model performance in terms of user engagement.
We used the ratio of the \textit{share}/\textit{dislike} frequency of top 5 ranked news to the overall \textit{share}/\textit{dislike} frequency in the dataset to measure \textit{share}/\textit{dislike} based performance, and we also reported the average finishing ratio of top 5 ranked news and their average dwell time if clicked.
We independently repeated each experiment 5 times and reported the average results.

\begin{table}[t]
	\centering
	%\resizebox{0.98\textwidth}{!}{ 
	\caption{Performance comparison in terms of news clicks. }\label{table.result}

\begin{tabular}{lcccc}
\Xhline{1.5pt} 
Methods & AUC             & MRR             & nDCG@5          & HR@5            \\ \hline
EBNR~\cite{okura2017embedding}    & 0.6112          & 0.2622          & 0.2790          & 0.1062          \\
DKN~\cite{wang2018dkn}     & 0.6076          & 0.2591          & 0.2768          & 0.1045          \\
NPA~\cite{wu2019npa}     & 0.6210          & 0.2685          & 0.2882          & 0.1095          \\
NAML~\cite{wu2019}    & 0.6192          & 0.2670          & 0.2871          & 0.1089          \\
LSTUR~\cite{an2019neural}   & 0.6224          & 0.2701          & 0.2896          & 0.1099          \\
NRMS~\cite{wu2019nrms}    & 0.6231          & 0.2707          & 0.2904          & 0.1103          \\
FIM~\cite{wang2020fine}     & 0.6250          & 0.2729          & 0.2925          & 0.1114          \\ \hline
DFN~\cite{xie2020deep}     & 0.6296          & 0.2748          & 0.2948          & 0.1140          \\
CPRS~\cite{wu2020cprs}    & 0.6334          & 0.2781          & 0.2972          & 0.1156          \\ \hline
FeedRec & \textbf{0.6609} & \textbf{0.3026} & \textbf{0.3304} & \textbf{0.1328} \\ \Xhline{1.5pt}
\end{tabular}
%}

\end{table}

\begin{table}[!t]
	\centering
	\caption{Performance comparison in terms of user engagement. $\uparrow$ Means higher is better, while $\downarrow$ means lower is better.}\label{table.result2}

	\resizebox{0.98\linewidth}{!}{ 
\begin{tabular}{lcccc}
\Xhline{1.5pt} 
Methods & Share($\uparrow$)           & Dislike($\downarrow$)         & Finish($\uparrow$)          & Dwell Time/s($\uparrow$)       \\ \hline
EBNR~\cite{okura2017embedding}    & 1.1203          & 0.9679          & 0.0671          & 84.061          \\
DKN~\cite{wang2018dkn}     & 1.1169          & 0.9729          & 0.0655          & 83.494          \\
NPA~\cite{wu2019npa}     & 1.1288          & 0.9588          & 0.0691          & 84.579          \\
NAML~\cite{wu2019}    & 1.1269          & 0.9593          & 0.0689          & 84.487          \\
LSTUR~\cite{an2019neural}   & 1.1325          & 0.9610          & 0.0696          & 84.712          \\
NRMS~\cite{wu2019nrms}    & 1.1343          & 0.9583          & 0.0709          & 84.793          \\
FIM~\cite{wang2020fine}     & 1.1365          & 0.9595          & 0.0711          & 85.010          \\ \hline
DFN~\cite{xie2020deep}     & 1.1398          & 0.9519          & 0.0745          & 85.346          \\
CPRS~\cite{wu2020cprs}    & 1.1455          & 0.9434          & 0.0772          & 86.129          \\ \hline
FeedRec & \textbf{1.2603} & \textbf{0.9011} & \textbf{0.0940} & \textbf{87.989} \\  
 \Xhline{1.5pt}
\end{tabular}
}
\end{table}

\subsection{Performance Evaluation}

First, we compare the performance of our \textit{FeedRec} approach with many baseline methods, including:
(1) EBNR~\cite{okura2017embedding}, an embedding-based news recommendation method with GRU network;
(2) DKN~\cite{wang2018dkn}, deep knowledge network for news recommendation;
(3) NPA~\cite{wu2019npa}, a neural news recommendation method with personalized attention;
(4) NAML~\cite{wu2019}, a neural news recommendation method with attentive multi-view learning;
(5) LSTUR~\cite{an2019neural}, a news recommendation method that  models long- and short-term user interests;
(6) NRMS~\cite{wu2019nrms}, using multi-head self-attention for news and user modeling;
(7) FIM~\cite{wang2020fine}, a fine-grained interest matching approach for news recommendation;
(8) DFN~\cite{xie2020deep}, deep feedback network for feed recommendation;
(9) CPRS~\cite{wu2020cprs}, a news recommendation approach with click preference and reading satisfaction.
The click-based and user-engagement performance of different methods are shown in  Tables~\ref{table.result} and \ref{table.result2}, respectively.
We have several findings from the results.
First, compared with the methods based on click feedbacks only, the methods that consider other user feedbacks (i.e., \textit{DFN}, \textit{CPRS} and \textit{FeedRec}) achieve better performance in terms of news clicks and user engagement.
It shows that \textit{click} feedbacks may not be sufficient to model user interests accurately and other feedbacks such as \textit{dislike} and dwell time can provide complementary information for user modeling.
Second, among the methods that can exploit multiple kinds of user feedbacks, \textit{CPRS} and \textit{FeedRec} perform better than \textit{DFN}.
This may be because the \textit{dislike} feedbacks are relatively sparse, which may be insufficient to distill negative user interests accurately.
Third, our \textit{FeedRec} approach outperforms other compared methods in both click- and engagement-based metrics.
This is probably because our approach can effectively exploit the various feedbacks of users to model their interests more accurately.
In addition, our multi-feedback model training framework not only considers news clicks but also the  engagement signals, which can help learn a user engagement-aware recommendation model to improve user experience.

\begin{figure}[!t]
	\centering
 
	\includegraphics[width=0.32\textwidth]{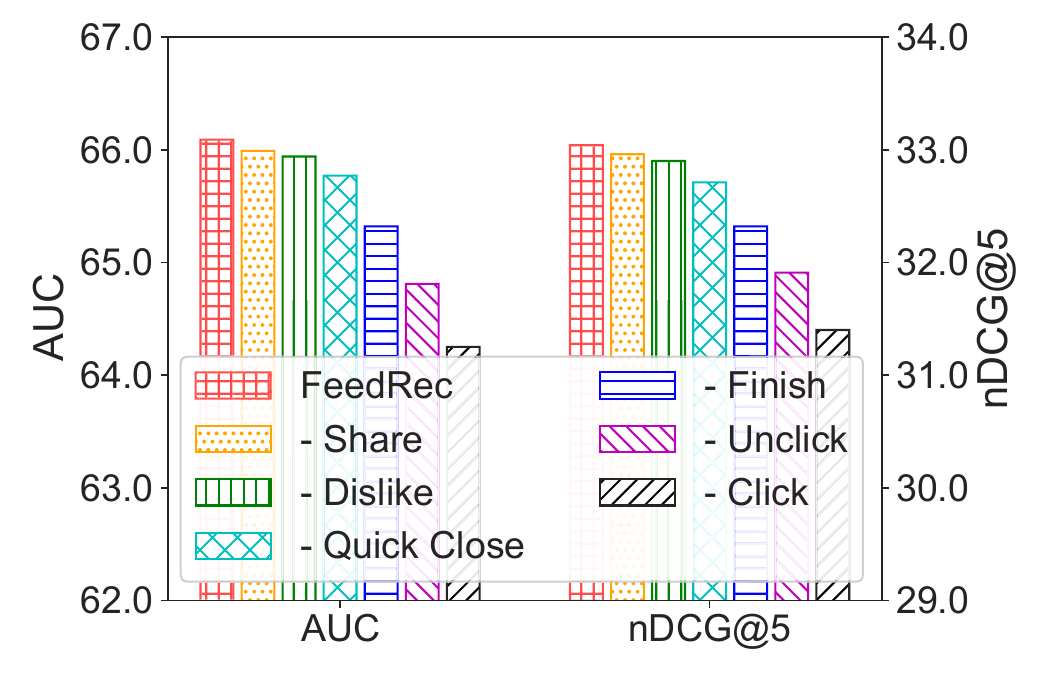} 
\caption{Influence of different types of user feedbacks.}\label{feed}  
\end{figure}

\begin{figure}[!t]
	\centering
 
	\includegraphics[width=0.32\textwidth]{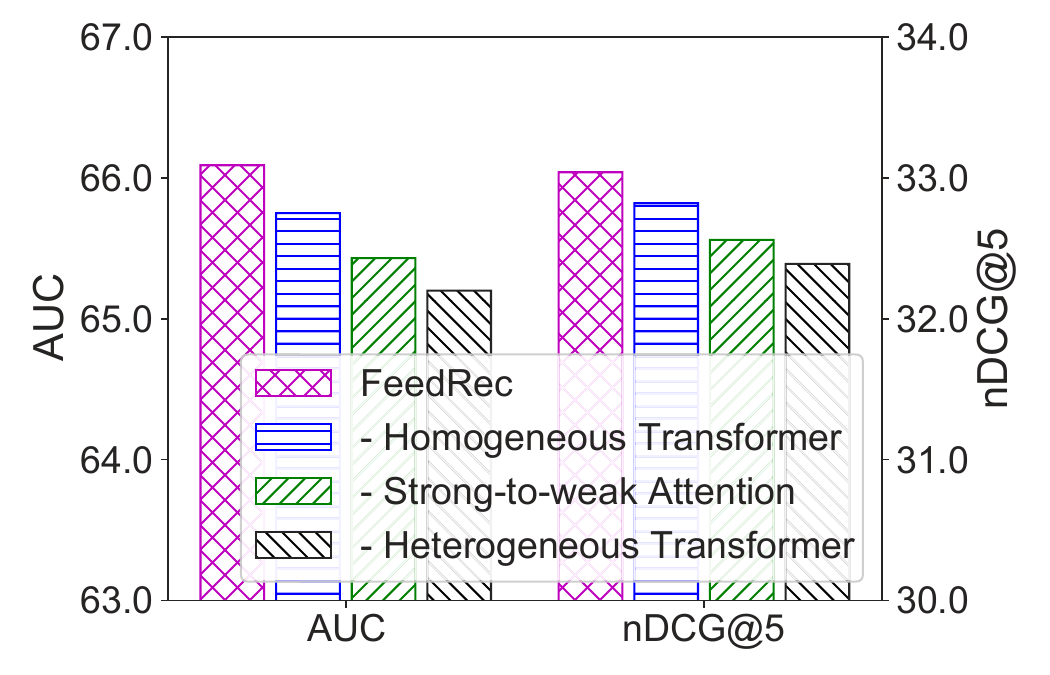} 
\caption{Effectiveness of several core model components.}\label{archi}  
\end{figure}

\begin{figure}[!t]
	\centering
 
	\includegraphics[width=0.32\textwidth]{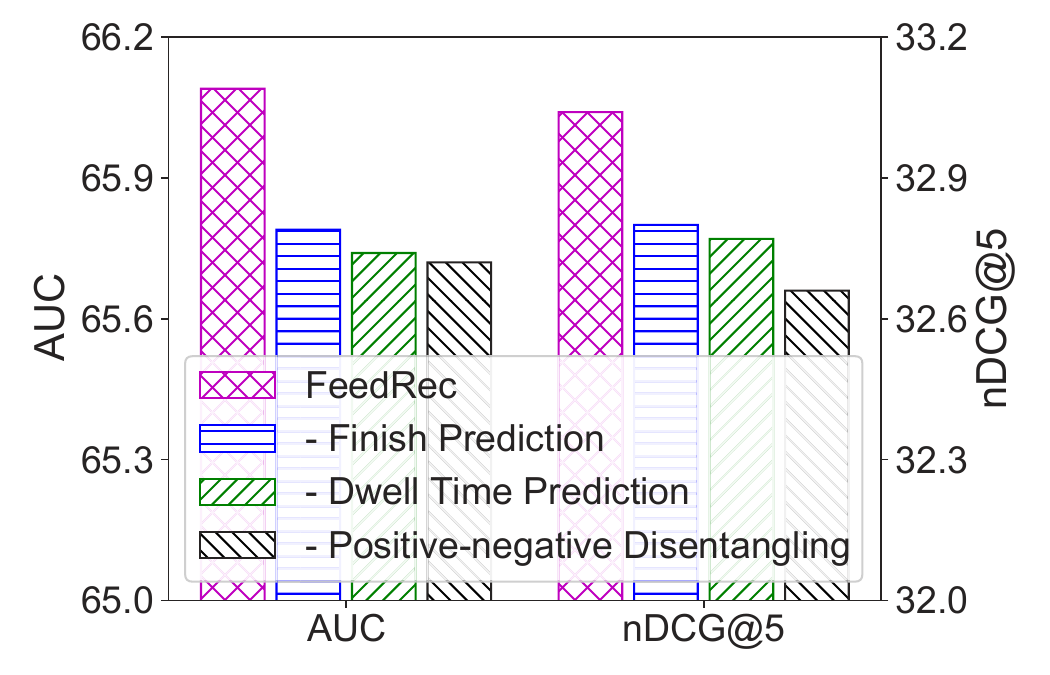} 
\caption{Influence of different loss functions.}\label{loss}  
\end{figure}

\begin{figure}[!t]
	\centering
 
	\includegraphics[width=0.32\textwidth]{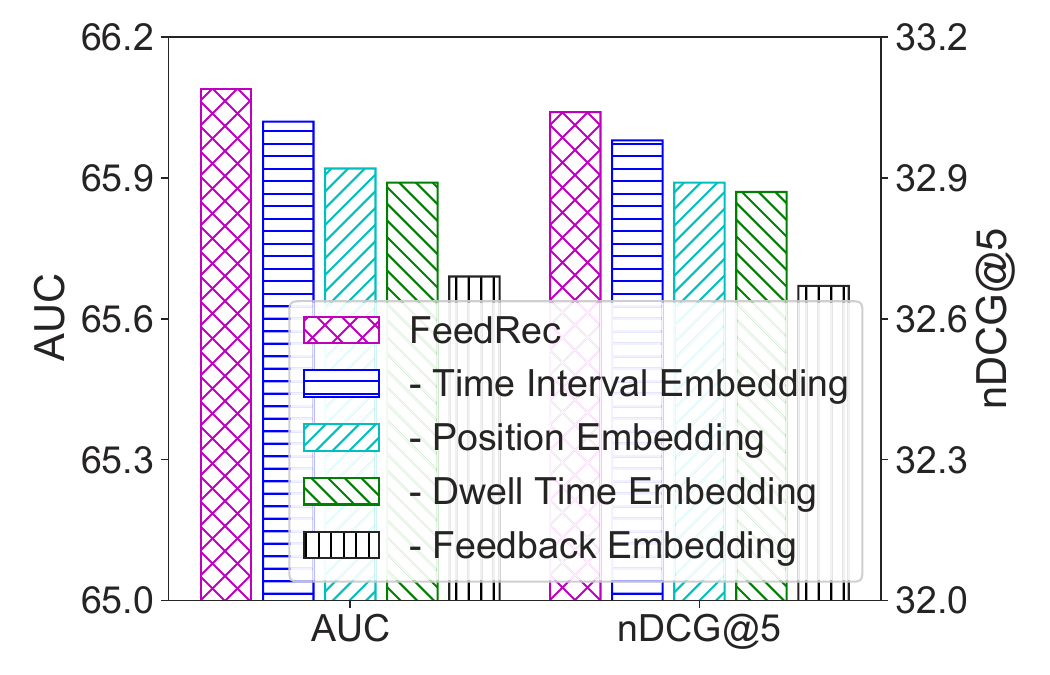} 
\caption{Effect of different embeddings in news encoder.}\label{emb}  
\end{figure}

\begin{figure}[!t]
	\centering
 
	\includegraphics[width=0.33\textwidth]{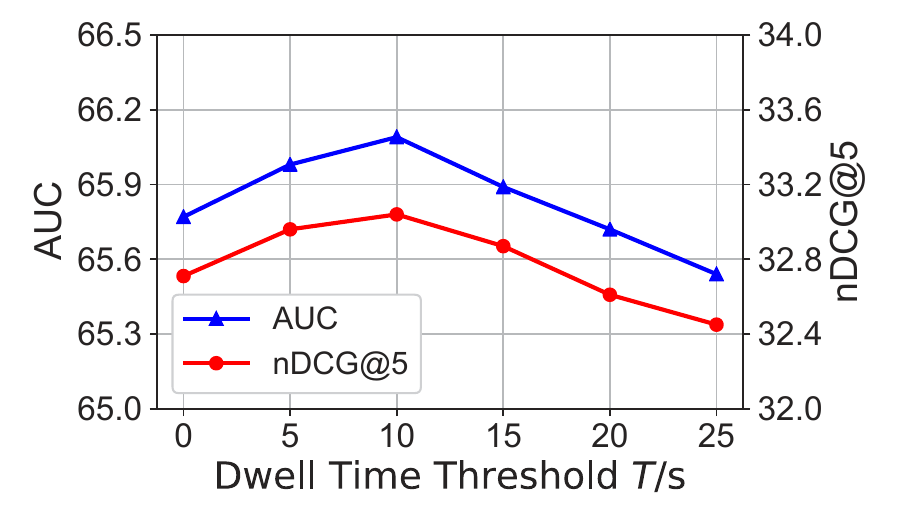} 
\caption{Influence of the dwell time threshold $T$.}\label{thre}  
\end{figure}

\begin{figure*}[!t]
	\centering

\subfigure[Finish prediction loss coefficient $\alpha$.]{\label{alpha}
	\includegraphics[width=0.28\textwidth]{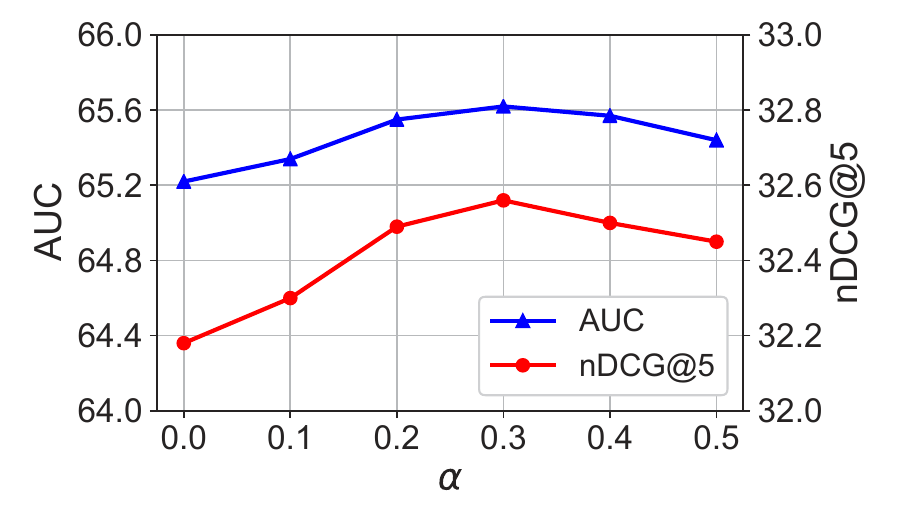}
	}
	\subfigure[Dwell time prediction loss  coefficient $\beta$.]{\label{beta}
	\includegraphics[width=0.28\textwidth]{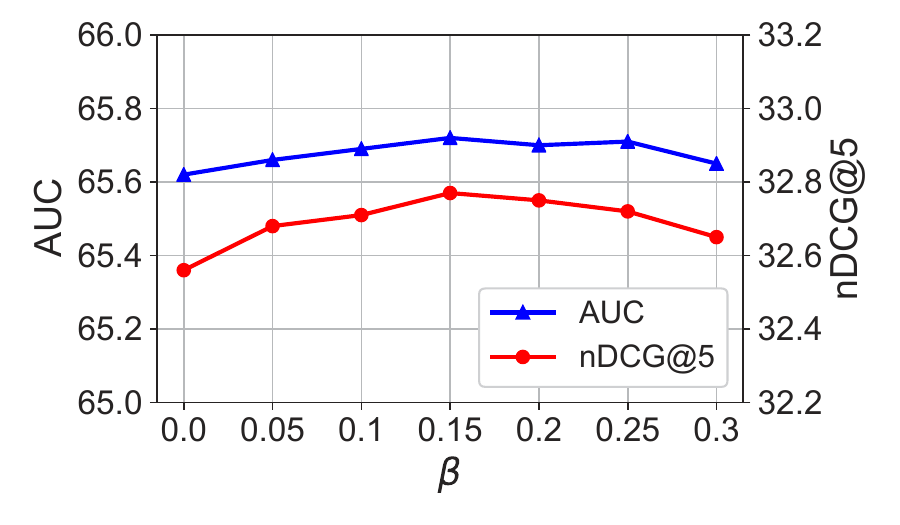}
	}  
	\subfigure[Positive-negative disentangling loss $\gamma$.]{\label{gamma}
	\includegraphics[width=0.28\textwidth]{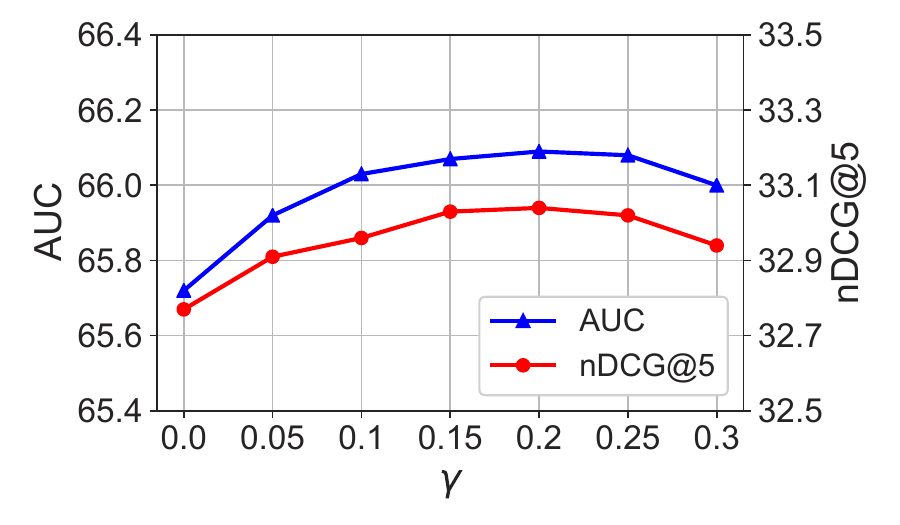}
	}  
\caption{Influence of different loss coefficients on the model performance.}\label{hyper}  
\end{figure*}
\subsection{Influence of Different Feedbacks}

Next, we study the influence of different feedbacks on the model performance.
We compare the performance of \textit{FeedRec} and its variants with one kind of feedbacks removed, and 
the results are shown in Fig.~\ref{feed}.
We find that the performance declines when any kind of feedbacks is dropped.
Among them, the \textit{click} feedback plays the most important role, which is intuitive.
However, we find it is interesting that the \textit{skip} feedback is the second most important.
This may be because skips can also provide rich clues for inferring user interests (usually negative ones) to support user modeling.
In addition, \textit{finish} and \textit{quick close} feedbacks are also important.
This may be because both kinds of feedbacks are indications of users' news reading satisfaction, which are important for modeling user preferences.
Besides, \textit{share} and \textit{dislike} feedbacks are also useful, but their contributions are relatively small.
This may be because that although these explicit feedbacks are strong indications of user preference, they are usually sparse in practice.
Thus, it is important to incorporate other implicit feedbacks like finish to model user interests more comprehensively.

\subsection{Model Effectiveness}

Then, we validate the effectiveness of the core model components in our \textit{FeedRec} approach and the loss functions used for model training.
We first compare the performance of our approach and its variants with one component removed, as shown in Fig.~\ref{archi}.
From the results, we find that the heterogeneous feedback Transformer contributes most.
This may be because the heterogeneous feedback Transformer can capture the global relatedness between the feedbacks of a user.
In addition, the strong-to-weak attention network is also very useful.
This is because it can select informative feedbacks for user modeling and meanwhile take the information of strong feedbacks into consideration, which can help distill  positive and negative user interests more precisely.
Moreover, the homogeneous Transformer can also improve the performance.
This may be because it can better capture the diverse characteristics of different kinds of feedbacks and benefit user modeling.

We also study the influence of each loss function on model training by removing it from the unified training loss. 
The results are shown in Fig.~\ref{loss}.
We find that the positive-negative disentangling loss can effectively improve the model performance.
This may be because it can push the model to distill  positive and negative interest information more accurately, which is beneficial for recommendation.
In addition, both the finish prediction  and dwell time prediction losses are helpful.
This may be because finish and dwell time signals are correlated to user satisfaction.
Thus, incorporating these signals into model training can  help learn an engagement-aware user model to improve the recommendation performance.

Finally, we investigate the influence of several different kinds of embeddings in the news encoder, including position embedding, feedback embedding, dwell time embedding and time interval embedding by removing one of them.\footnote{We do not report the scores without text embeddings because the performance is quite unsatisfactory.}
We illustrate the results in Fig.~\ref{emb}.
We find the feedback embedding plays the most important role.
This is because the embedding of feedback type is very useful for distinguishing different kinds of feedbacks.
In addition, the dwell time embedding is also important.
This may be because dwell time embeddings can provide rich information on inferring the satisfaction of users.
Besides, both position and time interval embeddings are  useful.
This is because position embeddings can help capture the feedback orders and time interval embeddings can help better model the relatedness between adjacent feedbacks.

\subsection{Hyperparameter Analysis}

In this section, we present some analysis on several critical hyperparameters in our approach, including the dwell time threshold $T$ for constructing \textit{quick close} feedbacks and the coefficients (i.e., $\alpha$, $\beta$ and $\gamma$) for controlling the importance of different tasks.
We first vary the threshold $T$ from 0 to 25 seconds to study its influence on model performance.
The results are shown in Fig.~\ref{thre}.
We find that the performance is suboptimal when the threshold $T$ is too small (e.g., 5 seconds).
This may be because many negative feedbacks with short reading dwell time cannot be exploited.
In addition, the performance also declines when $T$ goes too large.
This is because many positive feedbacks will be mistakenly regarded as negative ones, which is not beneficial for user interest modeling.
Thus, in our approach the threshold $T$ is set to 10 seconds, which is also consistent with the findings in~\cite{wu2020neural}.

We then study the influence of the three loss coefficients.
We first tune the finish prediction loss coefficient $\alpha$ under $\beta=\gamma=0$.
The results are shown in Fig.~\ref{alpha}.
We find that the performance is not optimal when $\alpha$ is either too small or too large.
This may be because the finish signals are not fully exploited when $\alpha$ is very small, while the main click prediction task will be influenced if the coefficient goes too large.
Thus, we empirically set $\alpha$ to 0.2.
Then, we tune the dwell time prediction loss coefficient $\beta$ under $\alpha=0.2$ and $\gamma=0$.
The results are shown in Fig.~\ref{beta}.
We find that there is also a peak on the performance curve.
This may be because the dwell time signals cannot be effectively captured if $\beta$ is too small, while the click prediction task is not fully respected when $\beta$ is too large.
Thus, we set $\beta$ to 0.15 according to the results.
Finally, we search the value of the positive-negative disentangling loss coefficient $\gamma$ under the previous settings of $\alpha$ and $\beta$.
We observe that a moderate value of $\gamma$ such as 0.2 is suitable for our approach.
This may be because the positive and negative feedbacks cannot be effectively distinguished when $\gamma$ is too small, while this regularization loss is over emphasized when $\gamma$ is too large.

\section{Conclusion}\label{sec:Conclusion}

In this paper, we present a general news feed recommendation approach that can exploit various kinds  of user feedbacks with different intensities.
In our approach, we propose a unified user modeling framework to incorporate various explicit and implicit user feedbacks to comprehensively capture user interests.
In addition, we propose a strong-to-weak attention network that uses strong feedbacks to distill accurate positive and negative user interests from weak implicit feedbacks.
Besides, we propose a multi-feedback model training framework to train the model in the click, finish and dwell time prediction tasks to learn engagement-aware feed recommendation models.
Extensive experiments on real-world dataset validate that our approach can effectively improve model performance in terms of both news clicks and user engagement.

\begin{acks}
This work was supported by the  National Key Research and Development Program of China under Grant No. 2018YFC1604000 / 2018YFC1604002.
\end{acks}
\bibliographystyle{ACM-Reference-Format}
\bibliography{main}

\end{document}